# BLOCK MEDCARE: ADVANCING HEALTHCARE THROUGH BLOCKCHAIN INTEGRATION

Oliver Simonoski[1], Dijana Capeska Bogatinoska[2]

[1]Faculty of Communication Networks and Security, University of Information Science and Technology, St. Paul the Apostle, Ohrid, 6000, Macedonia
[2]Faculty of Applied IT, Machine Intelligence and Robotics, University of Information Science and Technology, St. Paul the Apostle, Ohrid, 6000, Macedonia

## ABSTRACT

*In an era driven by information exchange, transparency and security hold crucial importance, particularly within the healthcare industry, where data integrity and confidentiality are paramount. This paper investigates the integration of blockchain technology in healthcare, focusing on its potential to revolutionize Electronic Health Records (EHR) management and data sharing. By leveraging Ethereum-based blockchain implementations and smart contracts, we propose a novel system that empowers patients to securely store and manage their medical data. Our research addresses critical challenges in implementing blockchain in healthcare, including scalability, user privacy, and regulatory compliance. We propose a solution that combines digital signatures, Role-Based Access Control, and a multi-layered architecture to enhance security and ensure controlled access. The system's key functions, including user registration, data append, and data retrieval, are facilitated through smart contracts, providing a secure and efficient mechanism for managing health information. To validate our approach, we developed a decentralized application (dApp) that demonstrates the practical implementation of our blockchain-based healthcare solution. The dApp incorporates user-friendly interfaces for patients, doctors, and administrators, showcasing the system's potential to streamline healthcare processes while maintaining data security and integrity. Additionally, we conducted a survey to gain insights into the perceived benefits and challenges of blockchain adoption in healthcare. The results indicate strong interest among healthcare professionals and IT experts, while also highlighting concerns about integration costs and technological complexity. Our findings underscore the transformative potential of blockchain technology in healthcare, pointing towards a new era of patient-centric and secure healthcare services. By addressing current limitations and exploring future enhancements, such as integration with IoT devices and AI-driven analytics, this research contributes to the ongoing evolution of secure, efficient, and interoperable healthcare systems.*

# 1. INTRODUCTION

The global healthcare sector faces stringent requirements for more efficient and accurate diagnostic processes, increasingly facilitated by digital healthcare systems. The adoption of Electronic Health Records (EHR) and established Health Information Exchange (HIE) systems has significantly reduced medical costs and improved the quality of healthcare [1]. Additionally, the integration of the Internet of Things (IoT) and smart healthcare systems enables remote communication with healthcare professionals and patient monitoring, enhancing overall well-being [2]. However, the digital transfer of highly sensitive medical data raises concerns about privacy and security, particularly in the context of targeted cyber-attacks [3].

Current systems often struggle with fragmented data storage, leading to inefficient data sharing among healthcare providers. This fragmentation poses a significant barrier to interoperability and effective communication, resulting in delays and potential errors in patient care. Moreover, these systems are vulnerable to data breaches due to their centralized nature, creating single points of failure that can be exploited by cyber attackers.

As traditional healthcare systems grapple with these challenges, a revolutionary technology has emerged as a potential solution: blockchain. This decentralized, immutable ledger technology has already disrupted various industries, including finance, supply chain management, and healthcare [4]. Beyond economic benefits, blockchain's impact extends to political, humanitarian, social, and scientific domains, offering promise for decentralized operations, improved governance, and enhanced collaboration.

This blockchain-integrated solution addresses the problems of fragmented data storage and security vulnerabilities in healthcare systems. It represents a significant advancement over traditional healthcare systems by tackling several critical issues inherent in existing approaches. Traditional solutions lack a unified framework that ensures both data security and seamless sharing among various stakeholders [5]. By leveraging blockchain's decentralized and immutable architecture, we create a comprehensive system that secures patient data while facilitating real-time access and updates for authorized healthcare providers. The consensus mechanism validates transactions across multiple nodes before recording them, preventing unauthorized data alterations and ensuring data integrity [6]. Cryptographic techniques, such as hashing, further safeguard sensitive medical information against unauthorized access and tampering.

Moreover, blockchain enables enhanced transparency and auditability [7]. Every transaction is recorded with a timestamp and a unique cryptographic signature, providing a complete and transparent history of data changes. This feature is particularly valuable in healthcare, where audit trails are essential for compliance and accountability. The blockchain-based approach also allows patients greater control over who accesses their information, fostering transparency and trust in the Healthcare process.

Unlike existing blockchain healthcare solutions that primarily focus on data security, our proposed framework uniquely integrates patient-centric care, optimized diagnostic processes, and enhanced data security within a single, comprehensive system. This holistic approach not only addresses current healthcare challenges but also lays the groundwork for a more interconnected and efficient healthcare ecosystem.

This research aims to overcome barriers to blockchain adoption in healthcare by proposing a blockchain-based framework that enhances patient-centric care, optimizes diagnostic accuracy, and strengthens data security across healthcare systems. By empowering patients with control over their medical records and fostering trust among stakeholders, our approach lays the groundwork for a more resilient and efficient healthcare infrastructure.

By implementing this blockchain-based framework, our research has the potential to revolutionize healthcare delivery, significantly reducing data breaches, enhancing diagnostic accuracy, and empowering patients. This could lead to improved health outcomes, reduced healthcare costs, and a fundamental shift towards more personalized and efficient healthcare systems globally.

The paper is organized into several sections, each focusing on essential aspects of the blockchain-integrated healthcare system. We begin with a comprehensive exploration of blockchain technology's theoretical foundations, followed by our Healthcare Development Methodology. We then detail the System Design and Architecture, Implementation and Testing processes, and present a critical evaluation of the project in our Results and Data section. Finally, we synthesize our research findings, offering insights into future work and recommendations for further enhancement of the developed system.

## 2. RELATED WORK AND CURRENT ADVANCES

The field of medical records management has seen significant evolution, with institutions transitioning from manual processes to digital solutions. However, both traditional and early digital methods face considerable challenges in ensuring data security, accessibility, and interoperability. This section examines recent advancements and persistent limitations in electronic health record (EHR) management systems.

In medical records management, institutions have historically relied on manual processes or web-based data management systems, both of which present significant vulnerabilities. These traditional methods, though aimed at making patient information accessible and organized, face notable limitations, including data handling challenges and heightened susceptibility to breaches. According to the authors of [1], the absence of a standardized and innovative framework has exacerbated these vulnerabilities, leading to disparate technologies that struggle with interoperability. The case study presented in publication [3] further emphasizes that, despite advancements in system-based approaches, there remains no consensus on the technical infrastructure required to effectively support patient-centered interoperability.

Recent efforts to address these limitations have explored various technological solutions, which can be broadly categorized into cloud-based and blockchain-based approaches:

*Cloud-based Solutions*:

Researchers [8] proposed a system architecture for managing EHRs using cloud services and granular access control. Their approach leveraged an intercloud infrastructure, which chains together multiple clouds, offering benefits such as end-to-end privacy and ease of data migration between providers. However, while this approach aimed to enhance scalability and security, it remains susceptible to exploits that could compromise patient data privacy.

*Blockchain-based Solutions*:

Another proposed framework [9] introduced the Action-EHR framework, built on Hyperledger Fabric, a permissioned blockchain platform. This system employs compliant cloud storage and advanced encryption methods to protect patient records. Despite its strengths in authentication and authorization, the Action-EHR framework faces challenges related to key management and the need for patient records to be uploaded multiple times for different healthcare providers, which can complicate the system's usability.

An additional approach discussed in recent research suggests enhancing the efficiency of EHR sharing in cloud environments through a hybrid blockchain architecture that combines public and private chains. This method aims to balance the security and privacy of sensitive health data while maintaining high availability and efficiency. However, the research also points out limitations, including limited real-world testing and potential privacy concerns associated with the public chain component [10].

In light of these existing approaches, our proposed system differentiates itself by addressing several critical gaps. While most studies focus on specific aspects of healthcare, such as data security or patient control, this research proposes a complete EHR system that integrates multiple healthcare functions into a single, blockchain-based framework. The approach not only incorporates role-based access control to enhance security but also introduces a user-friendly interface to address usability concerns, which have often been overlooked in previous studies.

Furthermore, our system addresses the common limitations identified in earlier research, such as scalability issues and high computational costs. In contrast to the intercloud approach, the system leverages blockchain's decentralized nature to further enhance data security and privacy while maintaining the flexibility needed for real-world implementation. Additionally, it addresses key management challenges by implementing a streamlined process that reduces the complexity for end-users, making the system more accessible to both patients and healthcare providers.

A unique aspect of our research is the inclusion of a user survey to gain insights into potential adoption challenges. By focusing on practical aspects such as user training, regulatory compliance, and real-world applicability, the designed system offers a more comprehensive and feasible solution for advancing the application of blockchain technology in healthcare.

Figure 1 below summarizes the key differences and advantages of our blockchain-integrated approach compared to traditional healthcare data management systems, highlighting the improvements in security, interoperability, and patient empowerment.

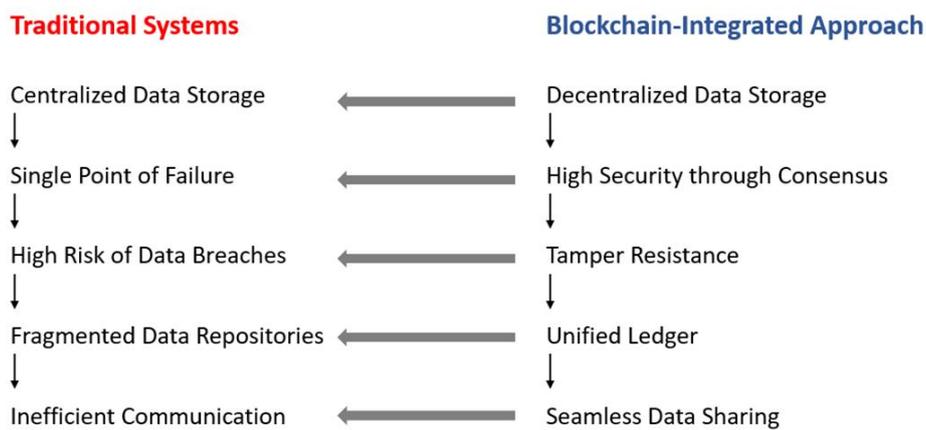

Figure 1. Traditional systems vs blockchain-integrated approach

In summary, while existing systems have made significant strides in addressing the storage and exchange of EHRs, our approach provides a more integrated and user-centric solution. By overcoming the limitations of previous systems and addressing practical challenges such as scalability and key management, our research contributes significantly to creating a more secure, efficient, and patient-centric healthcare infrastructure.

## 3. REVIEW AND THEORETICAL BACKGROUND

Blockchain technology has emerged as a powerful tool for enhancing data security, transparency, and decentralization across various industries, including healthcare. This section provides a comprehensive review of blockchain's fundamental principles, its evolution, and its specific applications in the healthcare sector. We begin by exploring the core concepts of blockchain, then delve into its classification and consensus mechanisms, and finally examine its architectural layers and implications for healthcare data management.

Blockchain is a distributed ledger technology that records transactions across a network of nodes, offering a decentralized approach that enhances security and reliability. Unlike centralized systems, which are vulnerable to a single point of failure, blockchain coordinates independent nodes that work collaboratively, ensuring the integrity and availability of data without the need for a centralized supervisor [11].

In the healthcare sector, a distributed model effectively balances security and accessibility, ensuring that patient data remains protected while allowing for continuous access and collaboration among healthcare providers. The evolution of blockchain technology, from its initial use in cryptocurrency (blockchain 1.0) to the introduction of smart contracts (blockchain 2.0), has now extended to healthcare with blockchain 3.0. Blockchain 3.0, refers to the next generation of blockchain technology that focuses on scalability, interoperability, and environmental sustainability. It encompasses features like advanced consensus algorithms, such as proof of stake and sharding, which enhance transaction speed and efficiency while reducing energy consumption, and enables seamless integration across different blockchain networks and industries. In the context of healthcare, blockchain 3.0 offers significant potential for improving patient data management, enhancing interoperability between healthcare providers, and ensuring the integrity and security of medical records.

Figure 2 illustrates the operational stages of blockchain, including transaction handling, block formation, and data synchronization across the network.

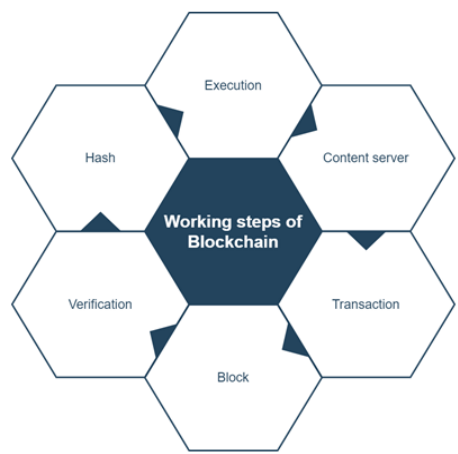

Figure 2. Operational steps of blockchain technology

In contrast to traditional e-health systems that rely on third-party intermediaries for authentication, blockchain's peer-to-peer data transfer reduces costs and ensures confidentiality. Blockchain employs asymmetric encryption methods, such as elliptic curve cryptography and RSA signature algorithms, to guarantee secure communication between nodes. Asymmetric encryption uses a pair of keys—one public and one private—to encrypt and decrypt data, allowing information to be securely transmitted so that only the intended recipient, who holds the private key, can access the content. The Secure Hash Algorithm (SHA) plays a crucial role in blockchain security, generating unique fingerprints for each block to ensure data integrity and prevent unauthorized modifications [12].

Blockchain technology is defined by three main features: security, transparency, and participation. Security is maintained through cryptographic techniques, transparency is achieved through publicly accessible ledgers, and participation is ensured by involving multiple nodes in the validation process. Key components of blockchain include platforms like Ethereum, smart contracts, the mining process for adding transactions, and consensus mechanisms for reaching agreement among nodes.

### 3.1. Classification of Blockchain and Consensus Mechanism

Blockchain systems can be categorized based on node selection methods, resulting in four types: public blockchains, consortium blockchains, private blockchains, and community blockchains [13]. Each type serves unique purposes with varying implications for security and decentralization. For instance, public blockchains are fully decentralized and accessible to everyone, while consortium blockchains feature pre-selected nodes and are often used by groups like financial institutions. Private blockchains provide restricted access, and community blockchains are used by organizations with specific use cases. In healthcare applications, the choice between these types depends on factors such as data privacy requirements, regulatory compliance, and the need for interoperability between different healthcare providers.

Consensus mechanisms are crucial in blockchain systems as they synchronize all nodes and ensure agreement on legitimate transactions for inclusion in the ledger. Several consensus algorithms are widely used:

- *Proof of Work (PoW)*: Nodes solve complex mathematical problems, with the first to solve creating the block and receiving a reward. While secure, PoW is resource-intensive and time-consuming.
- *Proof of Stake (PoS)*: Blocks are created by users who are randomly chosen based on their ownership stake, offering a more energy-efficient alternative to PoW.
- *Delegated Proof of Stake (DPoS)*: An extension of PoS, where elected trustees maintain the system and create blocks in a prescribed order, reducing transaction time.
- *Other Consensus Mechanisms*: These include Proof of Capacity (PoC) and Proof of Elapsed Time (PoET), which are less common and require further research to verify their effectiveness.

Table 1 summarizes key properties of consensus mechanisms, reflecting real-time data from the research and analysis stages.

Table 1. Comparison of Consensus Mechanisms

| Consensus mechanism | PoW | PoS | DPoS |
|---|---|---|---|
| Scenes | Public chain | Public chain Alliance chain | Public chain Alliance chain |
| Accounting nodes | All nodes | All nodes | Select representative nodes |
| Response time | ~ 10 minutes | ~ 1 minute | <1 minute |
| Ideal state Of Transaction Per Second | 7 TPS | 300 TPS | 500 TPS |
| Fault tolerance | 50% | 50% | 50% |

### 3.2. Blockchain Architecture and its Implications

Blockchain architecture is composed of five core layers: application, execution, semantic, propagation, and consensus [14]. Each layer plays a critical role in the functionality of the blockchain:

- *Application Layer*: Encodes end-user functionalities without relying on a centralized server.
- *Execution Layer*: Handles instructions from the application layer, including the execution of smart contracts.
- *Semantic Layer*: Validates transactions, establishes ownership rules, and defines data structures.
- *Propagation Layer*: Manages peer discovery and message transmission across the network.
- *Consensus Layer*: Ensures agreement among nodes on the state of the ledger, using various consensus methods like PoW to maintain system integrity.

Blockchain's core properties, immutability, decentralization, and verifiable information validity, make it especially suited for applications requiring secure, transparent, and tamper-proof data management. Its ability to facilitate cross-border transactions and maintain permanent records underscores its potential in scenarios involving multiple untrusted participants.

In conclusion, the unique features of blockchain technology, particularly in its latest iteration (blockchain 3.0), offer promising solutions to many challenges in healthcare data management. By providing a secure, transparent, and decentralized platform for storing and sharing medical records, blockchain has the potential to revolutionize healthcare delivery, improve patient outcomes, and enhance the overall efficiency of healthcare systems.

## 4. DEVELOPING HEALTHCARE SOLUTION: METHODOLOGICAL FRAMEWORK AND APPROACH

This section outlines the methodological framework and approach employed in developing our blockchain-based healthcare solution. We begin by describing the three-layer architecture that forms the foundation of our system, followed by a detailed examination of both functional and non-functional requirements. The section then delves into our research methodology, explaining the rationale behind our experimental design approach and the use of object-oriented analysis techniques.

## 4.1. System Architecture

The design of the proposed healthcare solution is structured into three distinct layers to ensure a comprehensive and efficient system. This three-layer architecture was chosen for its ability to separate concerns, enhance modularity, and improve scalability - crucial factors in developing a robust healthcare solution.

- *User Interface Layer*: This layer is responsible for displaying data and receiving user input, with interfaces designed to be compatible with both mobile and desktop devices, ensuring broad accessibility.
- *Business Logic Layer*: This critical layer facilitates the interaction between users and the blockchain-based Electronic Health Records (EHRs) platform. It ensures consistent data interfaces and standards by encapsulating user data into virtual transactions and resources. This data is then transferred to blockchain nodes, and new user data is stored within the blockchain database.
- *Data Access Layer*: This layer manages transaction verification, block creation, and the consensus process for voting on new blocks, ensuring that all nodes agree on validated transactions. It updates the blockchain database and maintains a customized blockchain containing all verified EHRs within the system.

Together, these layers create a robust and integrated solution that enhances system efficiency and data management. The seamless compatibility across mobile and desktop devices ensures accessibility for a wider user base, contributing to improved user experiences and broader acceptance.

## 4.2. System Requirements

Building upon this architectural framework, we identified several key requirements essential for developing a secure and efficient electronic health records system. These requirements were developed through the design science process described in the preceding section.

### 4.2.1. Functional Requirements

1. Implement a registration procedure for all entities;
2. Enable patient control over their records;
3. Allow doctors to upload records data;
4. Provide restricted access for specific entities to download data;
5. Implement encryption for each record or stored data.

Additionally, the system requires all users to access their accounts via MetaMask and authenticate themselves by logging into the system. Patients can access their medical records from anywhere at any time, provided they have a connection to the blockchain. The system supports appointment management, allowing patients to request appointments and doctors to retrieve these requests, update appointment details, and prescribe medication. Doctors can also input and send laboratory results directly to patients, ensuring a streamlined process for managing health information. All users have the capability to log out of the system, maintaining control over their sessions.

### 4.2.2. Non-Functional Requirements

- *Accessibility*: The application must be accessible via any browser that can connect to the blockchain, ensuring broad compatibility.
- *Usability*: The software is designed to be intuitive and straightforward, allowing users of all technological backgrounds to navigate it comfortably.
- *Security*: Each request within the system is assigned a unique transaction hash, providing verification and enhancing security.
- *Data Integrity*: The accuracy and verification of patient data are paramount, and the system ensures that all entered data is validated, maintaining the integrity of health information.

## 4.3. Research Methodology

This research employs an experimental design, systematically evaluating previous achievements to identify analogous solutions. The choice of experimental research design is motivated by its capacity for controlled testing and systematic analysis, which is essential given the limited number of prior implementations meeting the proposed system's requirements. This approach aligns with the study's objectives by enabling the exploration of various implementation strategies, ultimately leading to the development of a secure storage and exchange system for electronic health records that fulfills the specified requirements.

The study adopts an object-oriented analytic approach, utilizing techniques such as use-case modeling and sequence diagrams to clarify functional requirements and simulate system flows. This methodology introduces a patient control system that transforms traditional healthcare practices by consolidating all health data into a single, easily accessible platform.

Figure 3 illustrates the EHR System Architecture, depicting how data moves through each layer of our system. It highlights key processes such as user authentication, record access, and transaction verification, providing a comprehensive view of the system's operation.

This platform provides comprehensive access to medical history and examinations for authorized entities, streamlining data access for healthcare providers and patients alike. By centralizing health data management, our research significantly contributes to the field of blockchain in healthcare, addressing the growing demand for patient empowerment and data ownership. This innovative approach not only enhances patient control over their health records but also improves overall system efficiency and accessibility.

In conclusion, our methodological framework and approach, combining a three-layer architecture, comprehensive system requirements, and an experimental research design, provide a solid foundation for developing a blockchain-based healthcare solution that addresses key challenges in electronic health record management.

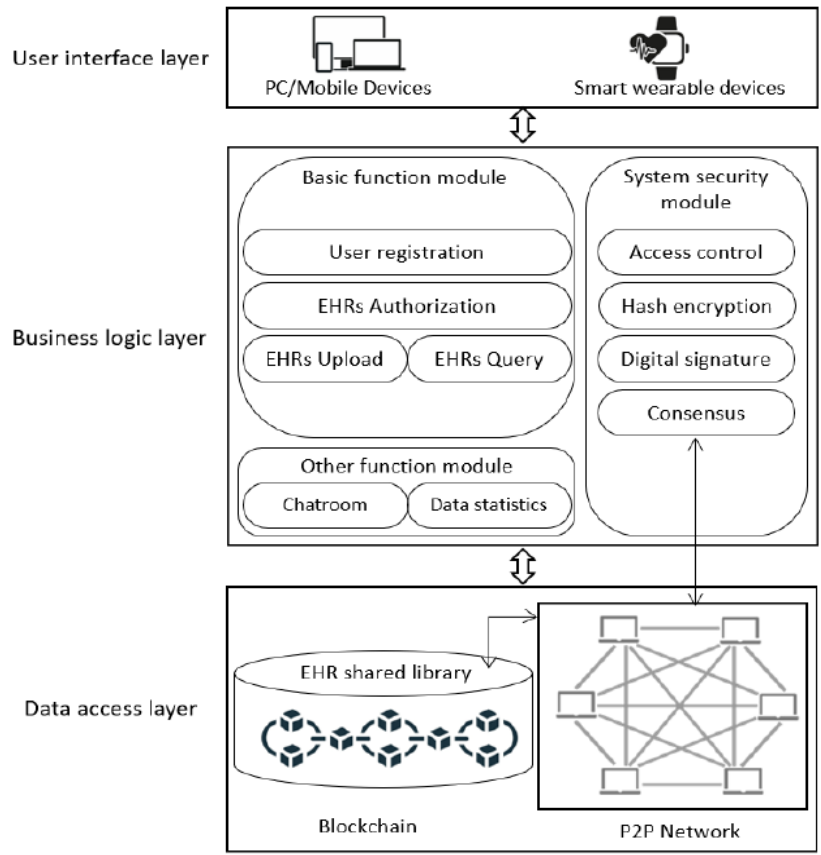

Figure 3. EHR System Architecture

## 5. SYSTEM DESIGN AND ARCHITECTURE

This section outlines the system design and architecture of our blockchain-based healthcare solution. We begin by describing the key entities involved and their roles within the system, followed by an explanation of the blockchain infrastructure and smart contract implementation. Finally, we detail the role-based access control (RBAC) mechanism that ensures secure and appropriate access to sensitive health data.

### 5.1. System Overview and Key Entities

The healthcare system is designed using a blockchain-based architecture involving three key entities: *patients*, *doctors*, and *administrators*. This approach was chosen to ensure data integrity, security, and interoperability. By implementing blockchain technology, we leverage advanced cryptographic techniques to secure medical records, ensuring that both patients and doctors can access records only with explicit patient consent, thereby enhancing data privacy and control.
The architecture is initiated by the administrator node, which is responsible for authentication, managing system variables, and enabling crucial functionalities. Patient and doctor nodes handle personal data management, appointment scheduling, and medical interactions. Figure 4 provides a detailed illustration of the entities, workflow, and operations of each entity within the system.

### 5.1.1. Entity Roles

*Administrator Node:*

The administrator node is depicted as the central authority, responsible for deploying smart contracts and initializing system operations. It authenticates users, manages medications, and sets system variables, ensuring the system's smooth functioning. This node also facilitates data exportation and oversees laboratory test parameters. Without the administrator node, other entities cannot operate effectively, as it provides essential authentication and configuration services.

*Patient Node:*

As shown in Figure 4, the patient node manages personal identification information and the patient's Ethereum address. Patients can update personal details and schedule appointments with specific doctors.

*Doctor Node:*

While not explicitly detailed in the provided text, we can infer that doctor nodes have functionalities related to accessing patient records, managing appointments, and inputting medical data.

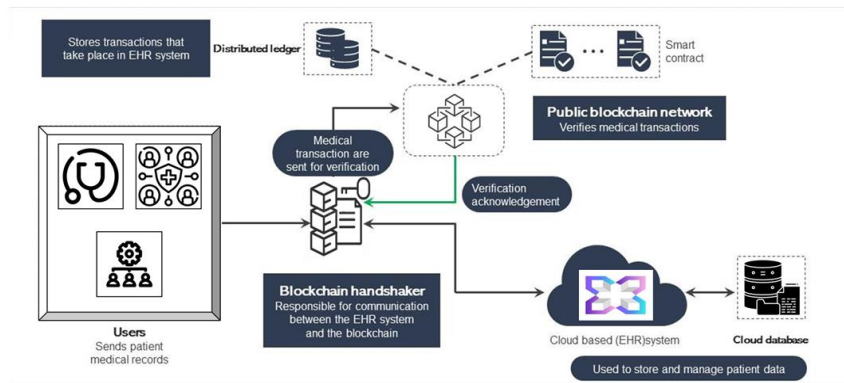

Figure 4. Architecture of proposed system model

## 5.2. Blockchain Infrastructure

The system employs a peer-to-peer (P2P) network structure that ensures optimal replication of Electronic Health Records (EHRs) data. In this decentralized network, each node has equal rights and obligations, which minimizes the need for extensive EHR storage on individual user devices and contributes to the system's overall validity and security. This P2P structure is particularly beneficial for EHR management as it enhances data availability, reduces single points of failure, and improves system resilience.

## 5.3. Smart Contracts

Central to the functionality of this blockchain-based system are smart contracts, which operate within the contract layer to facilitate secure and automated transactions between entities. These contracts enforce strict access controls, managing user permissions and ensuring that operations are executed only by authorized individuals. While the primary focus is healthcare, the application of smart contracts extends to other fields such as finance, supply chain management, and legal agreements.

The automation and security provided by smart contracts streamline various system operations, including:

- User login/registration
- Data management
- Appointment scheduling
- Prescription handling
- Laboratory results processing
- Data exportation

Administrative functions, such as adding medications and laboratory data, are restricted to users with administrator roles.

### 5.4. Role-Based Access Control (RBAC)

The EHR system employs role-based access control (RBAC), allowing administrators to define user permissions. Three default roles are established within the system: patients, doctors, and system administrators.

- *Patients* are permitted to perform actions related to their individual EHRs.
- *Doctors* have broader permissions, including the ability to add, query, and analyze EHRs, as well as engage in information exchange.
- *System administrators* hold the highest level of control, with the ability to modify user statuses, set system start dates, and adjust overall system settings.

Figure 5.a visually represents the RBAC structure, illustrating how the system determines user access based on their role.

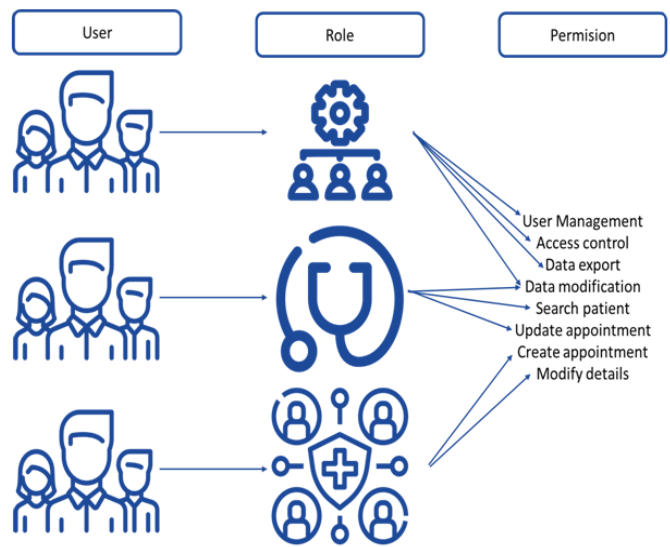

Figure 5. a) Role based EHR structure

The RBAC implementation follows the algorithm represented in Figure 5.b:

1. The system first checks if a user is signed in.
2. If not, the user is prompted to log in before further permission verification occurs.
3. Upon logging in, the system retrieves the user ID and validates permissions based on their role.
4. This role is obtained from the smart contract using the ID, ensuring that the user has the necessary permissions to access the requested resources.
5. If the user's operation permission matches the required authorization, access is granted; otherwise, access is denied.

This ensures that patients have restricted access to areas reserved for doctors and vice versa. The administrator has access only to the management site, where they define necessary operational data and manage user statuses (active or inactive).

These access control mechanisms ensure that the EHR system operates securely and efficiently, maintaining the integrity of medical records while enabling seamless interaction among users with different roles. By implementing these RBAC features, the system effectively manages permissions and access, aligning with the overall goal of creating a secure and user-friendly healthcare environment.

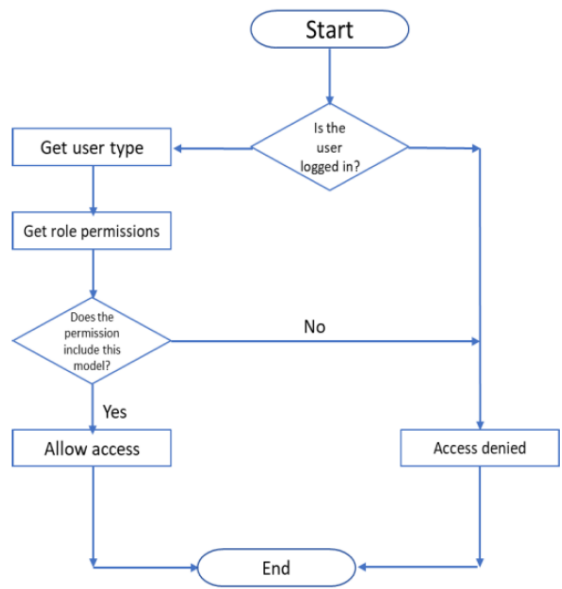

Figure 5. b) RBAC structure algorithm

In conclusion, the system design and architecture of our blockchain-based healthcare solution integrate advanced security measures, efficient data management, and user-friendly interfaces. The combination of blockchain technology, smart contracts, and role-based access control creates a robust framework for managing electronic health records while ensuring data privacy and integrity.

## 6. BLOCKCHAIN SOLUTIONS FOR HEALTHCARE: DESIGN, IMPLEMENTATION AND INSIGHTS

This chapter details the design, implementation, and evaluation of our blockchain-based healthcare solution. We begin by describing the development process and key technologies used, followed by an in-depth look at the user interface and functionality for different user roles. Finally, we present the results of a user survey conducted to assess the usability and potential adoption of the system.

In the dynamic realm of healthcare, the Electronic Health Records (EHR) system has undergone thorough validation procedures to ensure accuracy, reliability, and compliance with industry standards [15]. Despite these validation efforts, the need for strong security and data integrity persists. In this context, blockchain technology emerges as a transformative solution, providing an immutable ledger for all transactions within the EHR system. By utilizing blockchain technology, patient records are safeguarded against tampering and unauthorized access, thereby enhancing the security and trustworthiness of healthcare data.

### 6.1. Designing and Deploying Blockchain Solutions for Healthcare

#### 6.1.1. Technology Stack and Development Process

Utilizing blockchain technology, our system enhances security, privacy, and integration, effectively addressing persistent concerns in healthcare data management. The decentralized application (dApp) structure effectively enhances healthcare data management by ensuring a secure and user-friendly interface, powered by Node.js and React.js (see Figure 6). These technologies were chosen for their robust performance, extensive community support, and compatibility with blockchain development.

The development of the dApp begins with setting up the production environment, which includes configuring both the back-end and front-end systems. This architecture integrates smart contracts written in Solidity, which are tested and deployed using the Truffle suite and Ganache. Truffle simplifies the process of writing and deploying smart contracts, while Ganache offers a personal Ethereum blockchain for testing, allowing developers to verify contract functionalities in a controlled environment before deploying them to the Ethereum Virtual Machine (EVM) [16]. This setup promotes patient authorization, data security, and modernized data access, aligning with blockchain principles. Additionally, the integration of MetaMask for cryptocurrency wallet services helps ensure secure transactions, further enhancing the system's reliability.

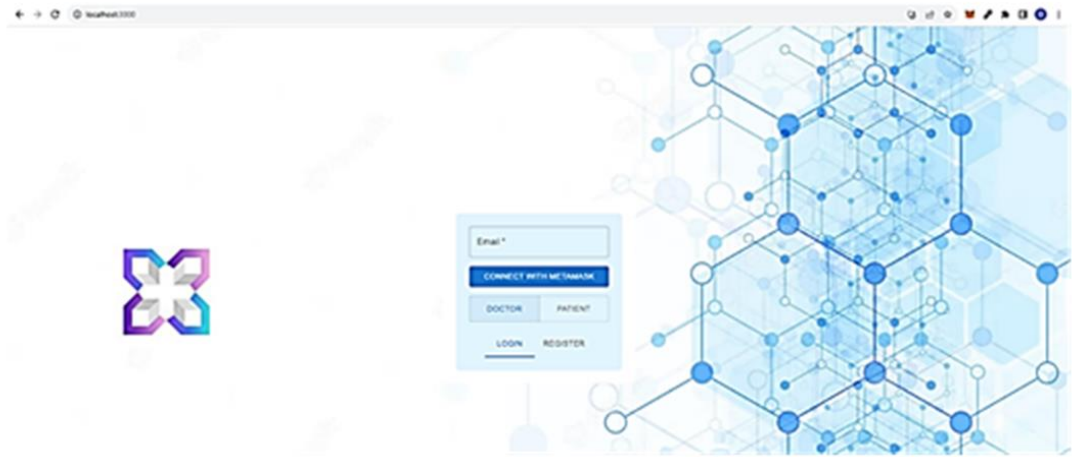

Figure 6. MedCare login/registration page

**6.1.2. User Interface Design and Functionality**

The system's interface prioritizes user convenience across different roles, as demonstrated in successful blockchain implementations in healthcare. For instance, a notable application of blockchain technology in healthcare involved its deployment within a major hospital network to manage patient consent for data sharing among various healthcare providers. This implementation not only enhanced patient control over data access but also led to a marked decrease in disputes related to consent and streamlined administrative processes within the hospital network [17].

*Doctors' Interface:*

The interface for doctors offers intuitive double-click interactions and a personalized dashboard for efficient appointment management (see Figures 7.a and 7.b). Key features include:

- Appointment rescheduling functionality with dynamic listing of available time slots
- Specialized tab for inputting and managing laboratory test results
- "E-reports" tab providing a comprehensive history of past activities

*Patients' Interface:*

The patient module emphasizes ease of use, transparency, and data integrity. Key features include:

- Effortless appointment scheduling with visual display of available time slots (Figure 8)
- Ability to track medical history and receive critical updates
- Structured presentation of laboratory test results

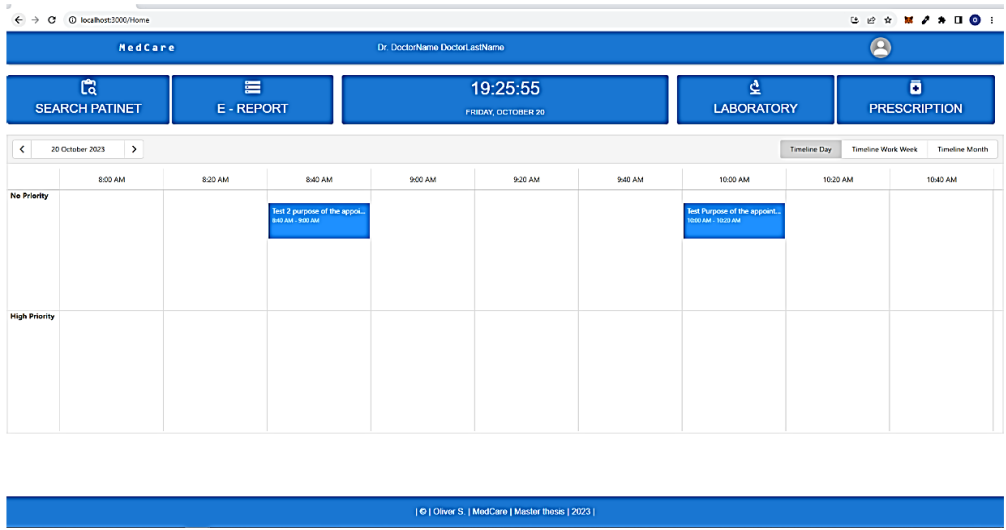

Figure 7.a) MedCare: Doctor dashboard

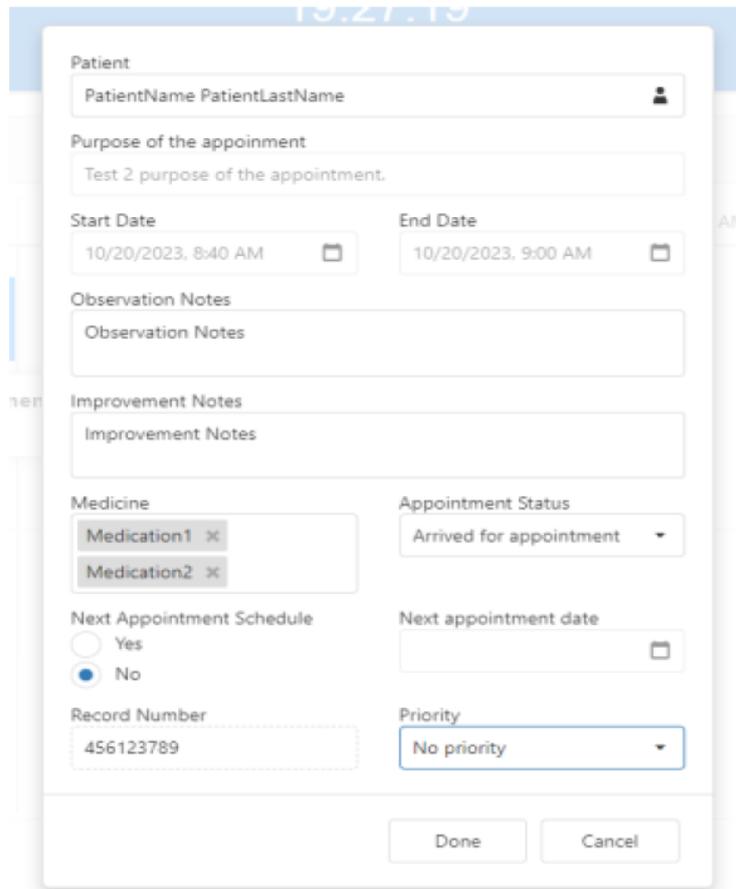

Figure 7.b) MedCare: Doctor appointment details

*Administrators' Interface*:

Administrators have access to specialized panels for comprehensive data management and oversight. Key features include:

- Monitoring and activation of registered users
- Management of medication inventories for doctor prescriptions
- Configuration of laboratory test parameters
- Data export capabilities in various formats (CSV, XML, TXT) for external analysis (Figures 9 and 10)

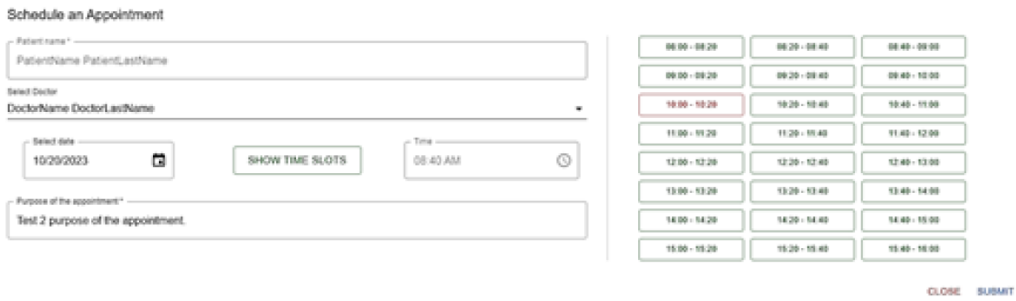

Figure 8. MedCare: Patient appointment scheduling

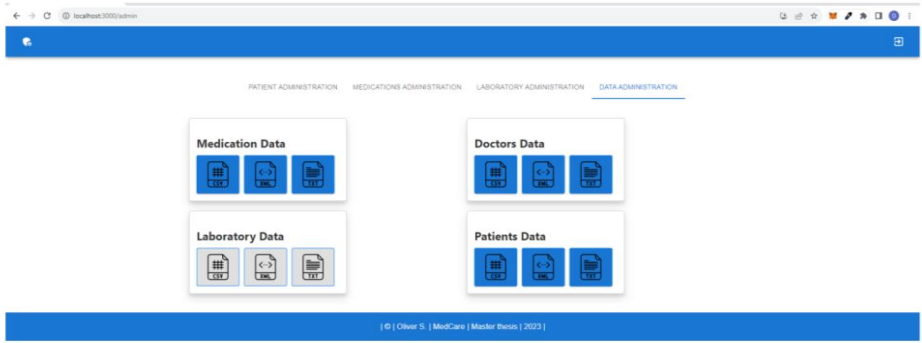

Figure 9. MedCare: Admin data administration page

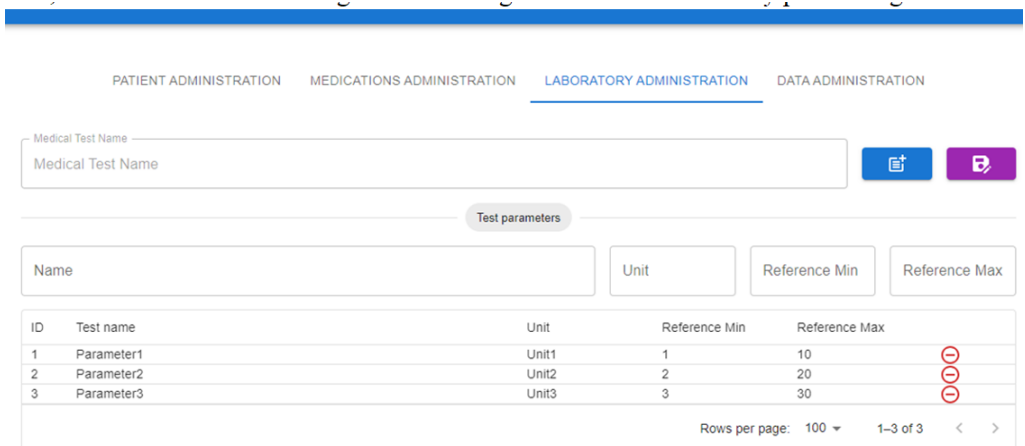

Figure 10. MedCare: Admin laboratory administration page

79

Another illustrative case involved a pilot program at a regional health authority where blockchain was used to track and verify the authenticity of pharmaceutical products. Each product's journey from manufacturer to pharmacy was recorded on the blockchain, creating a transparent and tamper-proof history of its handling. This approach reduced instances of counterfeit products entering the supply chain and increased trust among patients and healthcare providers [18].

### 6.1.3. Ethical Considerations

While implementing blockchain technology offers numerous benefits, it also raises important ethical considerations. Issues related to data ownership and patient autonomy must be carefully navigated to ensure that patients retain control over their health information. Additionally, the potential risks of technology misuse, such as unauthorized access or exploitation of sensitive data, necessitate robust ethical guidelines and governance structures. By prioritizing these ethical considerations, the implementation of blockchain can align with broader goals of promoting patient-centered care and protecting individual rights.

## 6.2. Testing and survey analysis: Ensuring reliability and exploring user perspectives

Having detailed the design and implementation of our blockchain solution, we now turn to the crucial aspects of testing and user feedback to ensure the system's reliability and usability.

### 6.2.1. Testing Methodology

The validation process ensures real-time accuracy in patient record updates, supporting continuous care. Testing methods play a crucial role in ensuring the reliability and functionality of the system, incorporating:

- Usability and functionality testing
- Measurement of specific metrics like response times and error rates
- Use of the Ethereum Virtual Machine (EVM) and tools like Ganache for local blockchain development

### 6.2.2. Survey Methodology and Results

To gain deeper insights into the usability and user-friendliness of the design, we conducted a survey targeting different groups of participants, including IT professionals and healthcare workers. The survey consisted of 28 questions covering various aspects related to design usability and user experience. Participants were selected to represent a diverse range of ages and occupations within the healthcare and IT sectors.

*Survey Results*:

- 66% of respondents reported moderate familiarity with blockchain technology, with only 13% indicating no familiarity at all (Figure 11.a).
- Responses regarding prior EHR system usage were evenly split, with 15 affirmatives and 15 negatives (Figure 11.b).
- Only 20% of respondents expressed concerns about the performance and scalability of blockchain technology in larger-scale applications (Figure 12.a).
- Most respondents (93.4%) believe in blockchain's potential to enhance interoperability and sharing across different sectors (Figure 12.b).

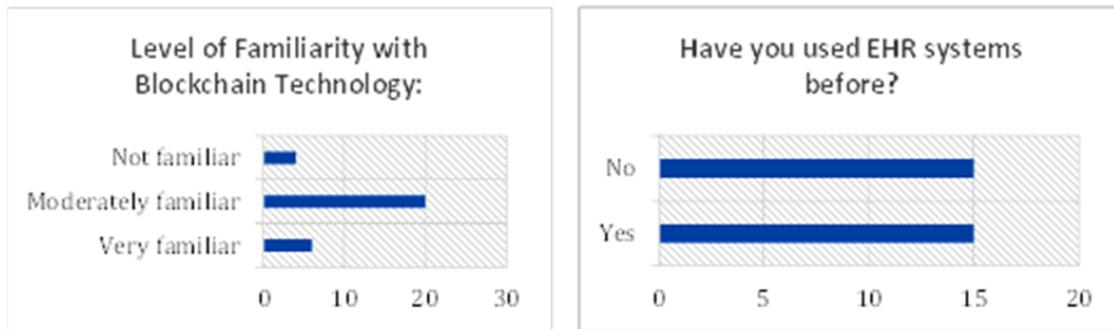

Figure 11.a. Results of a survey / Figure 11.b. Results of a survey

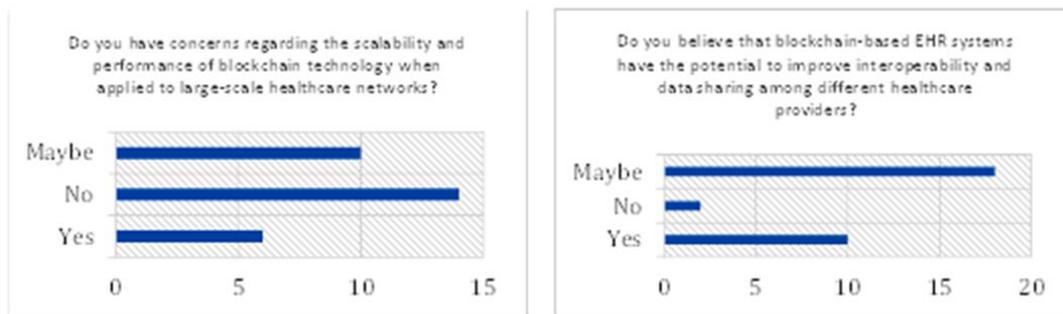

Figure 12. a): Participants' perspectives on the performance and scalability
b): Participants' perspectives on the potential of blockchain technology

*Key Findings*:

- Respondents noted several advantages over current systems, particularly in terms of data sharing among providers and paperwork reduction.
- They advocated for including blockchain education in courses while expressing concerns about integration costs and perceived technology complexity.
- Participants highlighted the critical roles of IoT and AI in integration, underscoring their importance in development and analytics.
- Overall, participants expressed strong interest in future discussions and research in this area.

In conclusion, our blockchain-based healthcare solution demonstrates promising potential for enhancing data security, patient control, and overall system efficiency. The positive survey results and user feedback provide encouragement for further development and implementation of blockchain technology in healthcare settings. Future work should focus on addressing the identified concerns, particularly regarding integration costs and education, to facilitate wider adoption of this innovative approach to healthcare data management.

## 7. CONCLUSION

This research investigates the integration of blockchain technology in healthcare, with a specific focus on enhancing the security and efficiency of data sharing. By addressing critical challenges associated with electronic health records, including confidentiality, scalability, user privacy, and the implementation of user-friendly cryptographic measures, this study contributes valuable insights into improving patient well-being and privacy through enhanced data security and scalability.

The primary focus of our research has been the integration of blockchain into healthcare systems to bolster security protocols, particularly for patient and doctor entities. Our proposed system emphasizes security through patient consent, authentication at each stage, and the use of public key signatures for data transactions, all validated by the blockchain. By leveraging cryptographic mechanisms and smart contract codes, we have created a secure environment that aligns with the goal of developing a security-centric healthcare system.

Survey results indicate a strong interest in adopting blockchain technology among participants, with many recognizing its potential to enhance data sharing and privacy. However, concerns were also raised regarding security, cost, and integration within existing healthcare infrastructures. Despite these challenges, the potential benefits of blockchain in healthcare are substantial. The proposed method's limitations, such as scalability and integration with current systems, suggest that a combined approach utilizing both traditional and blockchain technologies may be a practical first step, allowing for gradual adaptation and testing of blockchain's advantages.
For successful real-world deployment, several practical aspects must be addressed:

1. *Regulatory compliance*: Ensuring blockchain-based solutions meet stringent standards such as HIPAA or GDPR is essential for widespread adoption.
2. *Training*: Effectively engaging healthcare professionals in the use of new technologies is crucial to mitigate resistance to change and ensure smooth integration.
3. *Overcoming adoption barriers*: Addressing concerns such as cost, scalability, and interoperability with existing systems through pilot programs and collaborative efforts with industry stakeholders.
4. *Scalability*: Exploring blockchain 3.0 technologies, such as sharding and off-chain processing, to handle larger volumes of data efficiently.
5. *Integration costs*: Managing costs by leveraging existing infrastructure and gradually introducing blockchain components in a hybrid model.

Looking forward, future research should focus on:

1. Conducting experimental security tests to evaluate the system's resilience against potential attacks.
2. Exploring migration to and testing on blockchain 3.0 technology for improved performance.
3. Optimizing blockchain to address its inherent limitations.
4. Enhancing the application's functionality and user interface to ensure accessibility and ease of use.
5. Integrating Internet of Things (IoT) devices for real-time data monitoring and analysis.
6. Incorporating artificial intelligence (AI) to facilitate predictive analytics and personalized medicine.
7. Consolidating foundational components and offering succinct health data summaries through analytical tools.
8. Exploring health data forensics for improved security.
9. Establishing a messaging system aligned with health data exchange standards.

Furthermore, government-led initiatives could facilitate the adoption of blockchain technology in healthcare by establishing standardized frameworks and policies that promote interoperability and data sharing across different healthcare systems. Such initiatives can help create a unified national or regional healthcare network, reducing data duplication and enhancing the continuity of care.

In conclusion, while challenges remain, the integration of blockchain technology in healthcare offers promising solutions to longstanding issues of data security, privacy, and interoperability. By addressing the identified challenges and pursuing the outlined future directions, blockchain has the potential to revolutionize healthcare delivery, significantly improving patient outcomes and system efficiency. As we continue to refine and expand this technology, we move closer to a more secure, efficient, and patient-centric healthcare ecosystem.

**AUTHORS**

**Oliver Simonoski** received a Master's degree in Science in Information Technology from the University of Information Science and Technology "St. Paul the Apostle" in Ohrid, Macedonia. As a software developer, he focuses on innovation and efficiency in technology solutions, consistently integrating advanced approaches to enhance system performance and user experience. His main research interests include optimization and distributed computing, cloud technologies, IoT, blockchain technologies, and artificial intelligence.

**Dr. Dijana Capeska Bogatinoska** is an associate professor and researcher at the University of Information Sciences and Technologies "St. Apostle Paul" in Ohrid. With a background in electrical engineering and a doctorate in Applied Computer Science, she has a strong expertise in the field. Dr. Capeska Bogatinoska has participated in various national and international projects, including Horizon 2020, Erasmus+, and COST Actions. She has published over 60 papers in scientific conferences and journals and serves as a reviewer for reputable publications. Additionally, she is actively involved as a TCP member in multiple conferences and holds a position in the university senate.